# DEVELOPING MATHEMATICAL ORACLE FUNCTIONS FOR GROVER'S QUANTUM SEARCH ALGORITHM

**Cesar Borisovich Pronin**

**Andrey Vladimirovich Ostroukh**

MOSCOW AUTOMOBILE AND ROAD CONSTRUCTION STATE TECHNICAL UNIVERSITY (MADI)., 64, Leningradsky prospect, Moscow, Russia

**Abstract:** This article highlights some of the key operating principles of Grover's algorithm. These principles were used to develop a new oracle function, that illustrates the possibility of using Grover's algorithm for solving more realistic and specific search problems, like searching for a solution to a simple mathematical equation.

**Key words:** Grover's algorithm, oracle function, quantum informatics, qubit, superposition, quantum gate.

## Key operating principles of Grover's algorithm

**Grover's search algorithm (GSA)** is a quantum algorithm that can serve as an alternative to classical linear search algorithms. GSA can find solutions to a problem (function) by analyzing all possible states of a quantum register [2, 3].

The main part of the algorithm is the oracle function $f(x) = y$, which serves as a search criteria. It gets all of the possible quantum states of the register $x$ on input and inverts the amplitude value of states which can serve as its solutions. The complexity of the oracle function is not limited and it can serve all kinds of purposes.

Grover's algorithm can be composed of 3 main steps:
1) Forming uniform superpositions for all possible states of a quantum register, using Hadamard transform gates.

2) Developing and applying a specific oracle function, which will invert the amplitude of states, that serve as its solutions.

3) Applying an amplitude amplification function, which will amplify the amplitude of states that were chosen by the oracle.

Steps 2-3 are called **Grover iterations** and are repeated $N_G = \left\lfloor \frac{\pi}{4} * \sqrt{\frac{N}{l}} \right\rfloor$ times[1], to achieve the highest probability of receiving correct results after performing measurement. $N$ describes the number of all possible states of the quantum register $N = 2^n$, $n$ – number of qubits in the register, and $l$ is the number of expected solutions for the oracle function [3]. When the number of Grover iterations exceeds $N_G$, amplitude and probability values of states chosen by the oracle will start decreasing.

To find all solutions to a certain problem within all possible values of a register, in the worst case scenario, a classic linear search algorithm will be forced to check each possible value of the register, which will require $N = 2^n$, iterations. The difference in required iterations between Grover's algorithm and classical linear search can be calculated as $N_\Delta$ :

$$N_\Delta = \frac{N}{N_G} = \frac{N}{\left\lfloor \frac{\pi}{4} * \sqrt{\frac{N}{l}} \right\rfloor}$$

---

[1] $\lfloor x \rfloor$ – "x" rounded down to the nearest whole number

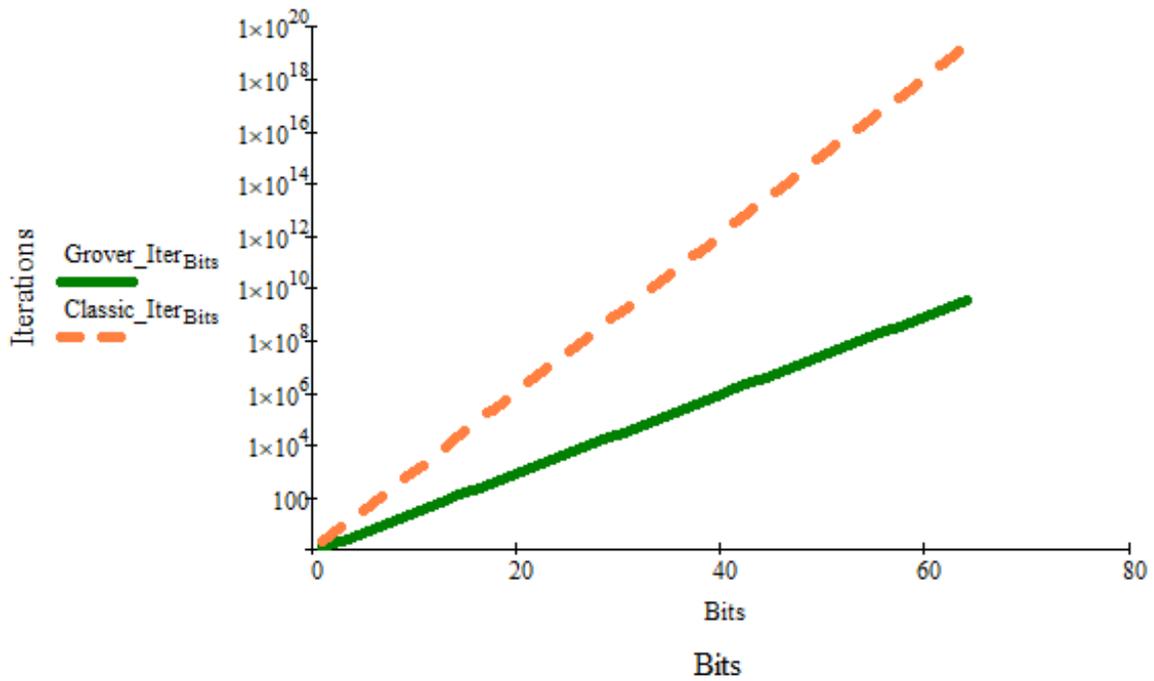

*Fig. 1. Logarithmic chart, that shows the difference between the number of required Grover iterations and linear search iterations, based on the number of bits in the analyzed register.*

Quantum circuit simulator «Quirk» is used for algorithm implementation and debugging purposes. It offers sensors, that allow to visualize transformations in quantum circuits [4].

Let's look at an example quantum circuit of Grover's algorithm, which was implemented in Quirk:

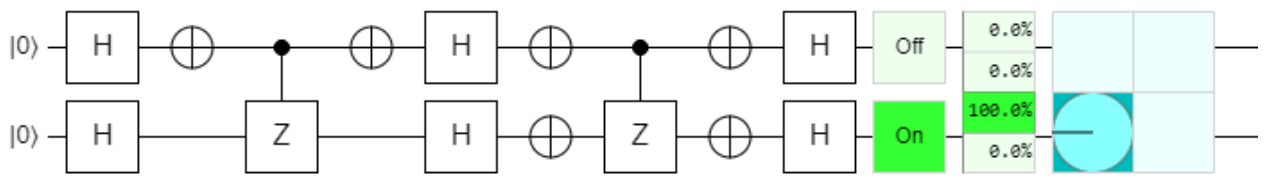

*Fig. 2. Example implementation of Grover's algorithm*

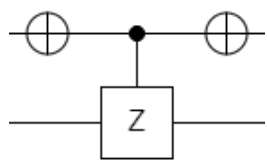

*Fig. 3. Oracle of this algorithm*

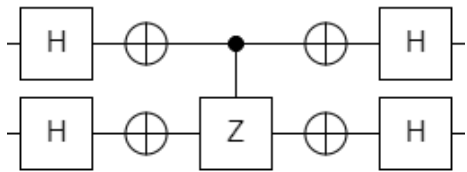

Fig. 4. Amplitude amplification

This circuit uses a 2-qubit register, so the number of required Grover iterations equals: $N_G = \left\lceil \frac{\pi}{4} * \sqrt{\frac{2^2}{1}} \right\rceil \approx \llbracket 1,57 \rrbracket = 1$

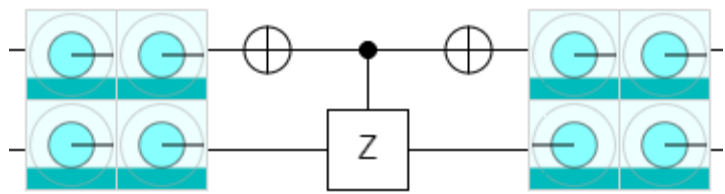

Fig. 5. Amplitude visualization shows that the example oracle function has inverted the amplitude of state $|10\rangle$

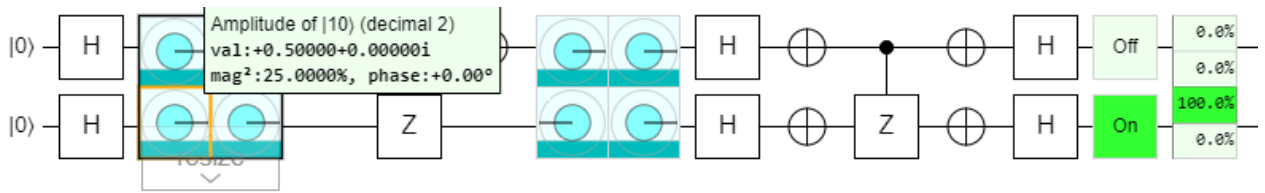

Fig. 6. Amplitude of $|10\rangle$ before applying the oracle function

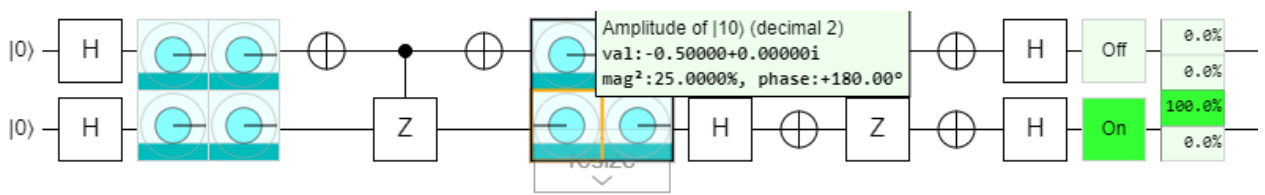

Fig. 7. Amplitude value of $|10\rangle$ after applying the oracle function

After the amplitude amplification step, the absolute value of state $|10\rangle$'s amplitude, increases.

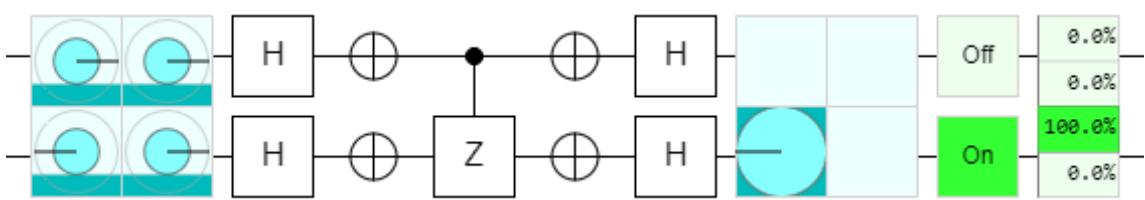

*Fig. 8. Amplitude amplification function and the results of running Grover's algorithm*

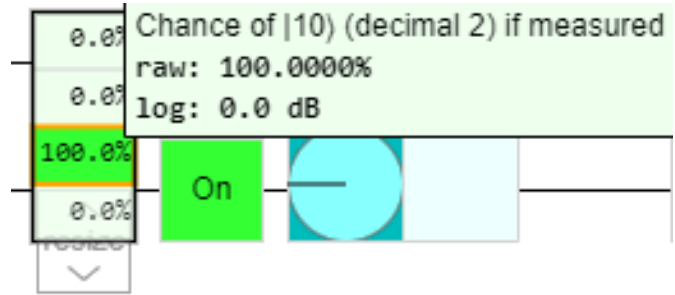

*Fig. 9. Final probability of state $|10\rangle$*

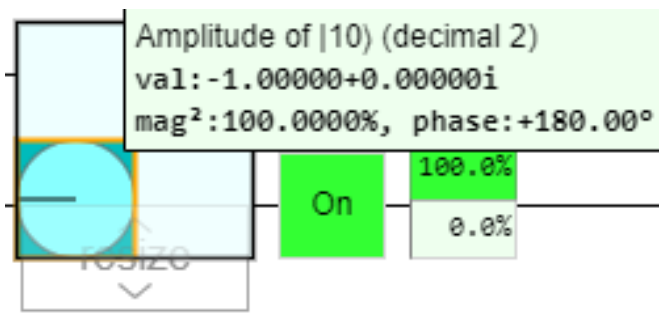

*Fig. 10. Final amplitude of state $|10\rangle$*

This is a common example of Grover's algorithm in which the goal of the Oracle function is to simply select and increase the probability of measuring the state $|10\rangle$. And since state $|10\rangle$ has the highest amplitude and superposition values, compared to other states, it becomes the most probable outcome of running this algorithm.

### Developing mathematical oracles for Grover's algorithm

To develop a more complex and sophisticated oracle function, it is often important to uncompute all operations, that are directly unrelated to the rotation of selected amplitudes. To achieve this, we make use of the reversibility property of quantum gates. We reverse all operations which are a part of the oracle, except the Controlled-Z rotation, as seen in fig. 11.

Using the concepts shown above, it is possible to make an algorithm with an oracle function for solving the problem of finding "x" in a simple equation $x + a = b$, without transforming it. To solve this problem having two bits of data (n

= 2), a classical linear search algorithm would require $N_c = 2^n = 2^2$ iterations, while Grover's algorithm would only need 1:

$$N_G = \left\lceil \frac{\pi}{4} * \sqrt{\frac{2^2}{1}} \right\rceil \approx 1{,}57 \approx 1$$

The oracle function for this problem is made using the arithmetic operators available in Quirk. For demonstration purposes we assigned values to a = 10 and b = 11. Circuit elements «a =», «b =», «X =» are labels, that are made for clarification purposes and do not have any effect on the circuit.

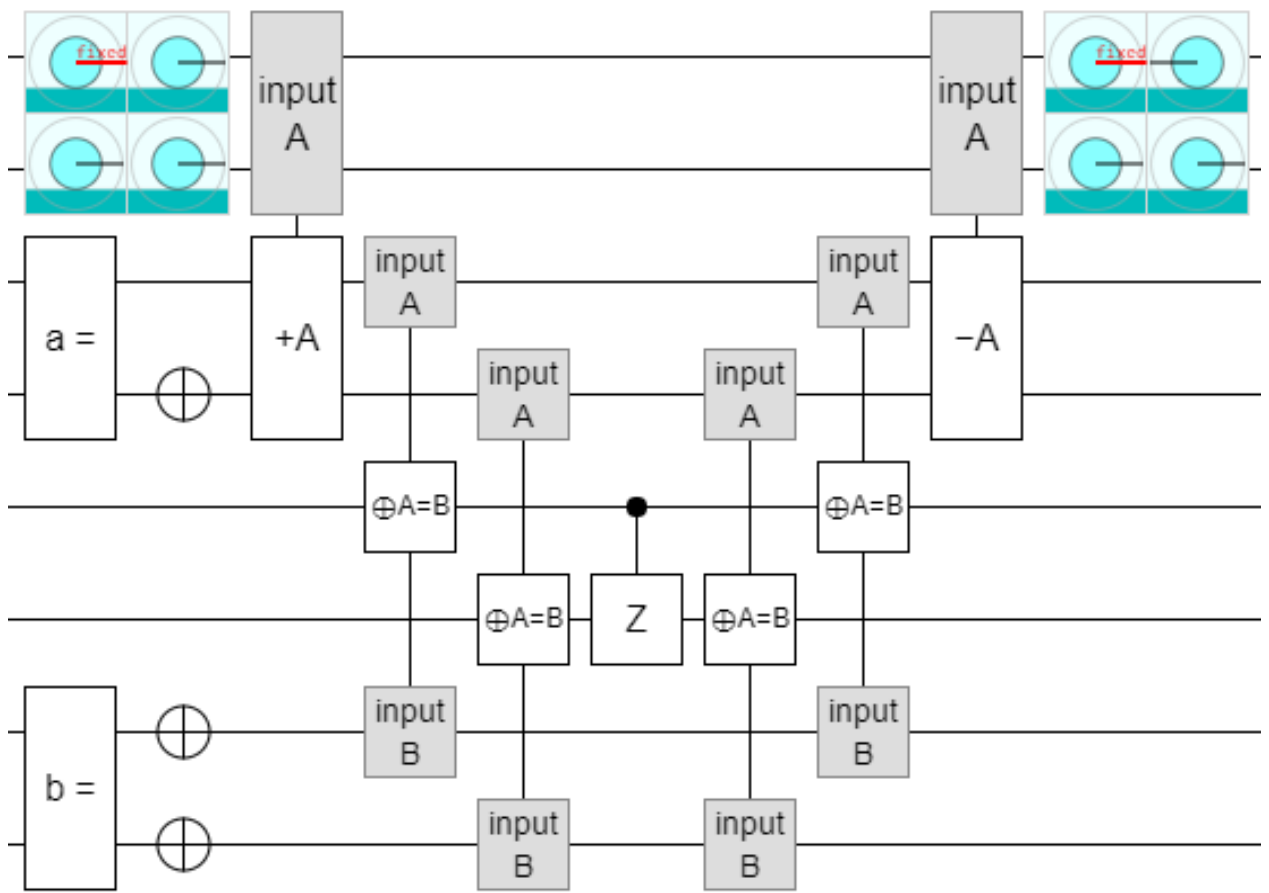

*Fig. 11. Oracle function for solving the example equation*

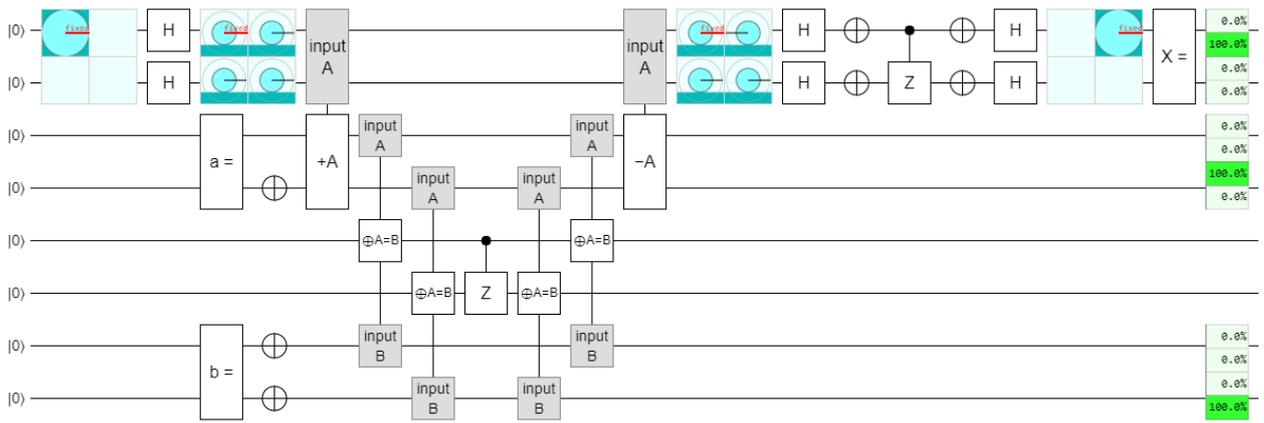

*Fig. 12. Grover's Search Algorithm solving the simple equation with result X = 01*

This solution is scalable, a 3 qubit version with a = 101 and b = 110 is shown on fig. 13-14.

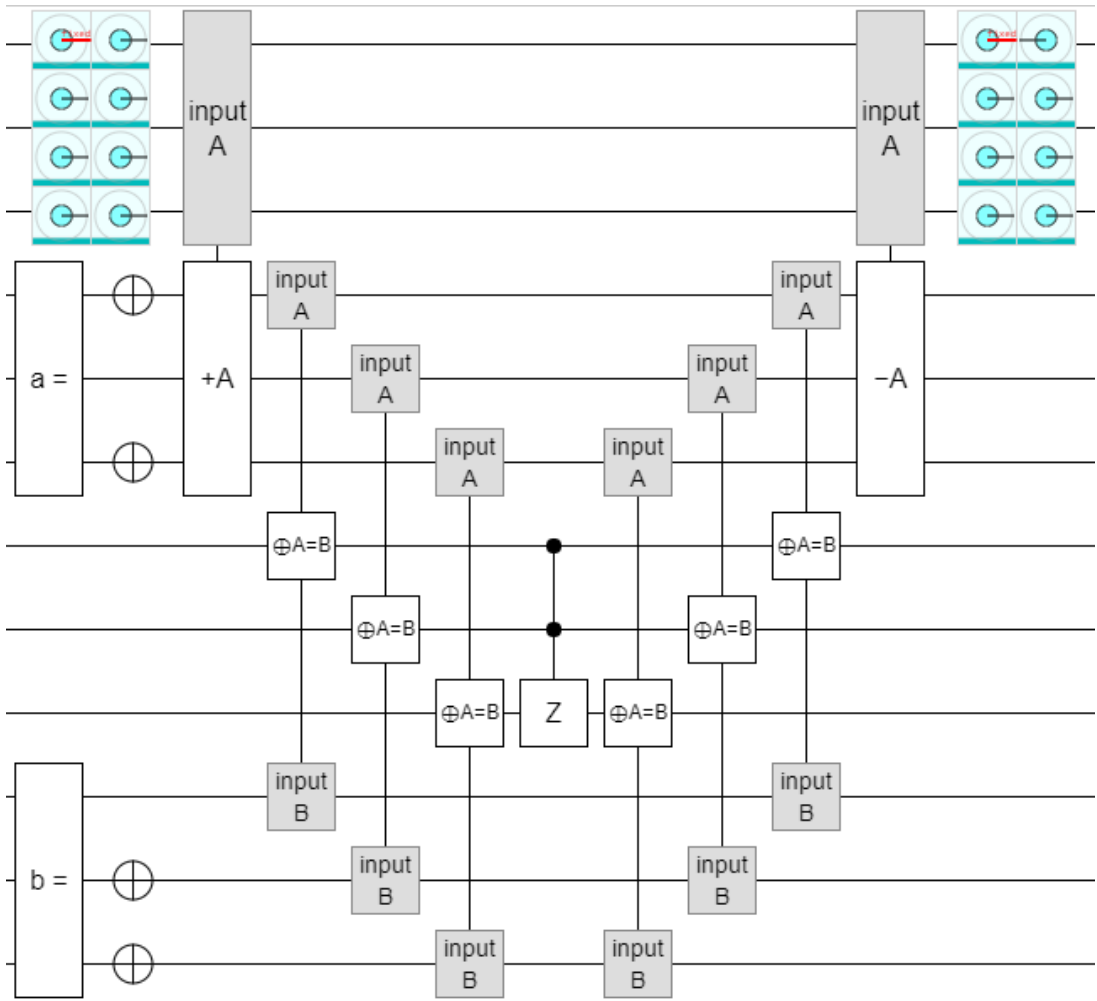

*Fig. 13. 3-qubit Oracle function for solving the example equation*

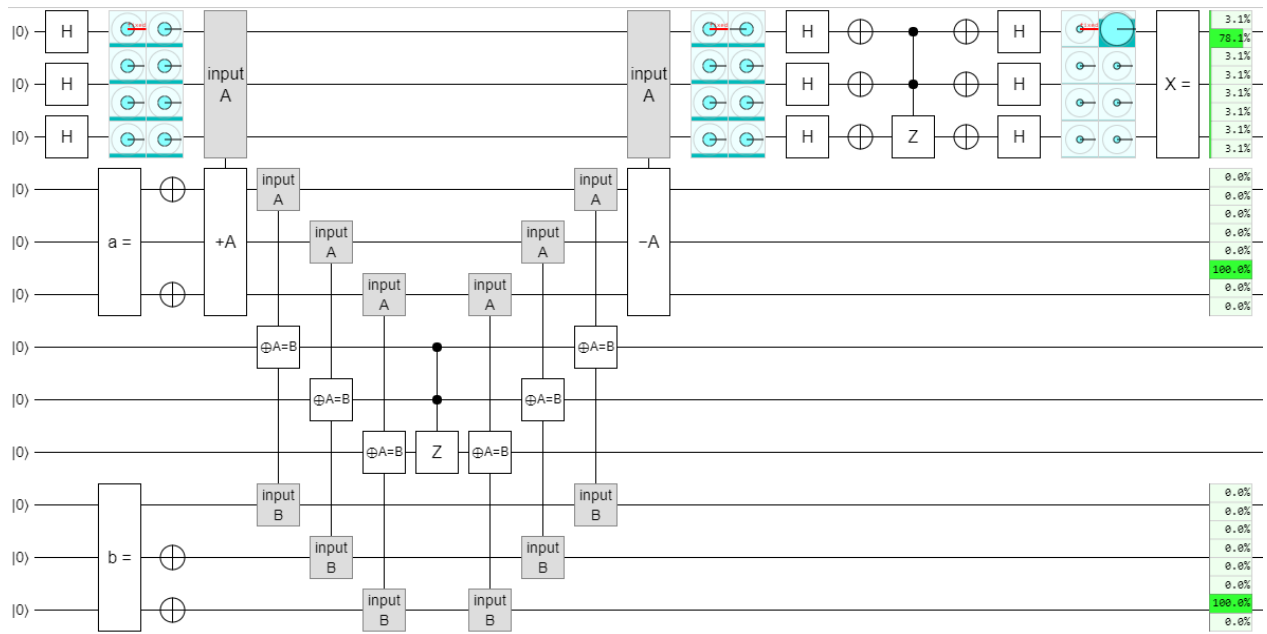

*Fig. 14. 3-qubit Grover's Search Algorithm solving the simple equation with result X = 001*

## Conclusions

This article has highlighted some of the key operating principles of Grover's algorithm, which were used to make a new oracle function, that could solve a simple and specific mathematical equation. This illustrates the possibility of using Grover's quantum search algorithm for solving more common and realistic search problems.

Grover's algorithm seems to be most efficient in problems that can't be effectively solved by common search algorithms due to high computational complexity, it is possible that in the future GSA could be applied to solve a large variety of problems that may be impossible or inefficient to solve with traditional computing methods.

**Author details**

**Andrey Vladimirovich Ostroukh**, Russian Federation, full member RAE, Doctor of Technical Sciences, Professor, Department «Automated Control Systems». State Technical University – MADI, 125319, Russian Federation, Moscow, Leningradsky prospekt, 64. Tel.: +7 (499) 151-64-12. http://www.madi.ru , email: ostroukh@mail.ru

**Cesar Borisovich Pronin**, Russian Federation, Postgraduate Student, Department «Automated Control Systems». State Technical University – MADI, 125319, Russian Federation, Moscow, Leningradsky prospekt, 64. Tel.: +7 (499) 151-64-12. http://www.madi.ru , email: caesarpr12@gmail.com